# Magnetically-Functionalized Self-Aligning Graphene Fillers for High-Efficiency Thermal Management Applications

J. Renteria[1,3,×] S. Legedza[2], R. Salgado[1,3], M.P. Balandin[2], S. Ramirez[3], M. Saadah[3], F. Kargar[1,3] and A.A. Balandin[1,3,*]

[1]Nano-Device Laboratory (NDL), Department of Electrical and Computer Engineering, University of California – Riverside, Riverside, California 92521 USA

[2]Quantum Seed LLC, 1190 Columbia Avenue, Riverside, CA 92507, USA

[3]Phonon Optimized Engineered Materials (POEM) Center, Bourns College of Engineering, University of California – Riverside, Riverside, California 92521 USA

[×] Affiliation during the time of this work: Quantum Seed LLC; Web: http://www.quantumseedllc.com

[*] Corresponding author (AAB): balandin@ece.ucr.edu ; Web: http://ndl.ee.ucr.edu

## Abstract

We report on heat conduction properties of thermal interface materials with *self-aligning* "magnetic graphene" fillers. Graphene enhanced nano-composites were synthesized by an inexpensive and scalable technique based on liquid-phase exfoliation. Functionalization of graphene and few-layer-graphene flakes with $Fe_3O_4$ nanoparticles allowed us to align the fillers in an external magnetic field during dispersion of the thermal paste to the connecting surfaces. The filler alignment results in a strong increase of the apparent thermal conductivity and thermal diffusivity through the layer of nano-composite inserted between two metallic surfaces. The self-aligning "magnetic graphene" fillers improve heat conduction in composites with both curing and non-curing matrix materials. The thermal conductivity enhancement with the oriented fillers is a factor of two larger than that with the random fillers even at the low ~1 wt. % of graphene loading. The real-life testing with computer chips demonstrated the temperature rise decrease by as much as $10^o$C with use of the non-curing thermal interface material with ~1 wt. % of the oriented fillers. Our proof-of-concept experiments suggest that the thermal interface materials with functionalized graphene and few-layer-graphene fillers, which can be oriented during the composite application to the surfaces, can lead to a new method of thermal management of advanced electronics.

### I.    Introduction

The increasing power densities in electronics made efficient heat removal a crucial issue for progress in information, communication and energy storage technologies [1-6]. Development of the next generations of integrated circuits (ICs) and ultra-fast high-power transistors depend on efficient heat removal [3, 4]. High-power-density devices such as Gallium Nitride (GaN) field-effect transistors (FETs) and GaN light-emitting diodes used in solid-state lighting require a better thermal management technology than is currently available [7-9]. Decreasing the temperature rise, $\Delta T$, in GaN transistors by only 10ºC doubles the life-time of the device while decreasing $\Delta T$ by 20ºC increases the transistor mean-time to failure by an order-of-magnitude [7]. A comparable reduction in the operating temperature of silicon (Si) complementary metal-oxide-semiconductor (CMOS) transistors would allow the chip manufacturers to substantially increase the clock speed of ICs. The demands for better thermal management are not limited to electronics. The power generation technologies for the photovoltaic solar cells also require efficient thermal management. Modern solar cells have an efficiency of ~15% in the conversion of light to electricity [10-12]. More than 70% of solar energy is lost as heat and has to be removed from the cell to prevent performance degradation [12-14].

The most important and commonly used component of passive thermal management is thermal interface material (TIM). There are different types of TIMs, including curing and non-curing thermal pastes, phase-change materials and solid heat spreaders. The function of TIM is to fill the voids and grooves created by the imperfect surface finish of two connecting surfaces and to improve surface contact and the conduction pathway across the interface. Typical TIM consists of a base (matrix) material and fillers, which are used to increase the overall thermal conductivity. Conventional fillers include silver, aluminum oxide and other metal or ceramic particles. Large loading fractions ($f$>50 vol. %) may be required in order to achieve desirable thermal conductivity. Development of more efficient TIMs is crucial for improving heat removal and reducing $\Delta T$ of a wide range of devices. The common strategy for improving TIM performance is finding the right filler material with the high intrinsic thermal conductivity, which can couple well with the matrix and attach to the connecting surfaces. While the thermal conductivity and thermal resistance of TIM are important characteristics of the material, the ultimate metric for the performance of TIMs is the temperature rise reduction, which can be achieved with it in a given device – heat sink assembly.

The discovery of unique heat conduction properties of graphene [15-18] promptly led to the proposals of the use of graphene and few-layer graphene (FLG) as fillers in TIMs [19-23]. In the thermal context, we consider a flake to be FLG rather than a piece of graphite as long as its thickness is below 7-10 atomic planes, and correspondingly, Raman spectrum is different from that of bulk graphite [24]. For practical thermal applications, FLG can have certain benefits as compared to single layer graphene. The thermal conductivity of FLG is still high and it is subject to less degradation when FLG flake is embedded inside matrix material as compared to that of graphene [18]. The larger cross-sectional area of FLG translates to higher heat fluxes along the length the flake as compared to single-layer graphene. Significant enhancement of the bulk thermal conductivity of epoxy with the addition of a proper mixture of graphene and FLG (with the loading $f$<10 vol. %) was reported in several studies [19-23]. The results with other matrix materials such as paraffin wax ($C_nH_{2n+2}$ hydrocarbons) were also promising [25, 26]. The above mentioned studies used a random uniform mixture of graphene and FLG fillers in the matrix. A uniform dispersion of the fillers and the absence of air bubbles are important for the improved heat conduction properties of the prepared composites [21-23, 26, 27].

The theory considerations for graphene composites [28, 29] and experimental results for other types of fillers suggest that a strong increase in thermal conductivity at small loading fraction $f$ can be achieved if the fillers are aligned along the direction of heat flux. In the TIM context the direction of alignment should be perpendicular to the connecting surfaces, thus, facilitating heat transfer from one surface (e.g. computer chip) to another (e.g. heat sick or package). Recent molecular dynamics (MD) simulations predicted that one can achieve a tremendous ×400 enhancement of the thermal conductivity along the direction of the graphene flake alignment at small $f$=5 vol. % in common matrix materials [28]. In the direction perpendicular to the alignment no thermal conductivity increase was obtained. The experimental results with other fillers, such as carbon nanotubes (CNTs) [30-37], are in line with this theoretical prediction. TIMs with low graphene and FLG filler loading ($f$<5 vol. %) are strongly preferable because high $f$ results in increased viscosity, air gap formation and agglomeration, which degrade heat conduction properties. The low loading of graphene is also beneficial for keeping the price of TIMs in the acceptable range.

In this paper, we show that functionalizing graphene and FLG with $Fe_3O_4$ nanoparticles can help one to achieve the goal of alignment of the fillers during the dispersion of the TIM. The proposed method is inexpensive and scalable for industrial use. The strongly enhanced heat conduction properties are evidenced from the measurements of the *apparent* thermal conductivity and temperature rise in actual heat generating devices. The testing with computer chips demonstrated that the temperature rise can decrease by as much as $10^oC$ with use of non-curing thermal interface material with ~1 wt. % of the oriented graphene fillers. Previous experiments with aligned fillers used CNTs grown by the chemical vapor deposition (CVD) or by the microwave plasma-enhanced chemical vapor deposition (PECVD) [30-37]. The approach based on CVD growth requires high temperature processing and complicated assembly. It is also prohibitively expensive for most of TIM applications. Our results indicate that TIMs with low loading of the functionalized graphene and FLG fillers, which can be oriented during the composite application to the surfaces, have the potential for a breakthrough in thermal management of advanced electronics.

## II.     Synthesis of the Graphene-Enhanced Thermal Interface Materials

The graphene and FLG solution was produced following the scalable liquid-phase exfoliation (LPE) method [38, 39]. The functionalization was achieved via the recipe previously developed for CNTs used in magnetic and biomedical applications [40, 41]. It involves poly-sodium-4-styrene-sulfonate (PSS) as a wrapping polymer and polyelectrolyte poly-dimethyl-diallylammonium chloride (PDDA) for a homogeneous distribution of positive charges [42-49]. In this approach, the positive charges ensure the adsorption of negatively charged magnetic nanoparticles onto the surface of graphene and FLG by means of electrostatic interactions. The process produced graphene fillers dressed with magnetic nanoparticles of ~10 nm average diameter. We also found that a mixture of graphene and FLG flakes with magnetic nanoparticles followed by temperature treatment under certain conditions likewise resulted in attachment of nanoparticles to graphene fillers without intermediate chemical processing steps.

The functionalization of graphene with magnetic nanoparticles followed the CNT route, which was demonstrated for applications other than thermal management [40-51]. The method combines polymer wrapping technique (PWT) and layer-by-layer (LBL) self-assembly allowing the non-covalent attachment of nanoparticles to the carbon filler leaving intact their structure and thermal properties (see Figure 1). The non-covalent bonding is an important aspect of the procedure because it preserves the intrinsically high thermal conductivity of graphene [18, 27]. Stronger covalent bonding sometimes used for increasing CNT filler – matrix coupling often results in defect formation leading to phonon scattering and TC reduction at least at some filler loading fractions [27]. We utilized poly-sodium-4-styrene-sulfonate (PSS) as a wrapping polymer providing stable dispersions of carbon fillers (both CNTs and graphene). Owing to the high density of sulfonate groups on the negatively charged polyelectrolyte PSS, the PSS coating acts as a primer on the graphene surface for subsequent homogeneous adsorption of the cationic polyelectrolyte poly-dimethyl-diallylammonium chloride (PDDA) through the electrostatic interactions [41, 52, 53]. The deposited PDDA layer, in its turn, provide a homogeneous distribution of positive charges. The positive charges ensure the efficient adsorption of negatively charged magnetic nanoparticles onto the surface of graphene by means of electrostatic interactions. The adsorption of nanoparticles (diameter range D~6-10 nm) on graphene surface is achieved easier than that on CNT surfaces due to CNT's high curvature, which hinders the formation of dense coatings. The magnetic nanoparticles prepared in solution (basic pH) are negatively charged and therefore are electrostatically attracted to the positively charged PDDA layer adsorbed on graphene fillers. It was reported for CNTs that the pH for the most efficient adsorption of $Fe_3O_4$-$\gamma$-$Fe_2O_3$ nanoparticles on polyelectrolyte was found to be 11.9-12.0 [41].

[Figure 1: Schematic of Functionalization]

The steps for preparing epoxy-based TIMs with "magnetic graphene" fillers were similar to the ones described by some of us elsewhere for regular LPE graphene [19] (see Figure 2). The epoxy based components were weighed with the intended loading wt% of LPE graphene powder to the resin (Epoxy-Mount Resin 145-10010) and the hardener (Epoxy-Mount Hardener 145-10015) at the manufacturer's 10:3 ratio guideline. The composites were evenly mixed (Flacktek DAC 150) under vacuum conditions. The vacuum pump accessory was used to prevent development of the bubbles trapped as a result of mixing dry materials with liquids. The optimized speed mixer setting was found to be approximately 500 rpm for 20 seconds. The low mixing speed and time were used owing to the high sheering of the dry graphene powder in the non-cured epoxy at higher mixing speeds (>1000 rpm) that caused the epoxy to cure faster than desired. Several cycles of mixing and vacuuming were applied to achieve the uniformly mixed composites. The composites were exposed to the magnetic (H=1.2 T) for the flake ordering. The synthesis of non-curing TIMs with "magnetic graphene" followed a similar process. We used commercial base (Loctite TCP 4000 D PSX-D). The base material was weighted and the desired graphene loading fraction was added using a speed mixer (Flacktek DAC 150) for uniform distribution. The mixing was performed from 300 rpm to 1000 rpm for 1 min to 5 min between incremental additions of graphene filler and base materials until the desired weight percent of graphene, viscosity and smoothness were achieved. The process was carried out under vacuum conditions to evacuate air bubbles from the composite. The resulting mixture should have a smooth texture. The roughness is an indication of excessive air mixed into the composite, which strongly degrade the thermal conductivity. The graphene enhanced TIMs were sandwiched between two surfaces of interest, e.g. thin copper (Cu) foils, for thermal testing. Two glass slides were used to apply even pressure to the sandwiched TIM between two connecting surfaces. The flake alignment was achieved by placing the surface-TIM-surface sandwich, e.g. Cu-TIM-Cu structure, on a permanent magnet (K&J Magnetics 1.2 Tesla NdFeB).

[Figure 2: TIM Preparation]

Figure 3 illustrates the structure and properties of the "magnetic graphene", i.e. graphene and few-layer-graphene flakes functionalized with magnetic nanoparticles. The scanning electron microscopy (SEM) and transmission electron microscopy (TEM) images have been used to confirm the attachment of magnetic nanoparticles (see Figure 3 (a-b)). For the proof-of-concept experiments we did not attempt to have only single-layer graphene and perfect dispersion. Some agglomeration was allowed and even desirable for visualization of flake alignment. We have previously demonstrated that a mixture of graphene and FLG works better for thermal management applications [19, 21, 20]. The functionalized graphene and FLG flakes were incorporated into the matrix materials, e.g. epoxies or non-curing natural oil bases. The alignment of the functionalized graphene flakes with an external magnetic field was made possible with the recent development of strong permanent magnets [54-55]. Such magnets, e.g. nickel plated neodymium (NdFeB), can provide the near-field magnetic field intensity in the range from $0.5 - 2$ T needed for the proposed application. It is important that the magnetic field lines in such magnets are predominantly perpendicular to the connecting surfaces.

[Figure 3: SEM, TEM, and Optical Image of Flakes Reacting to Magnetic Field]

### III. Thermal Conductivity and Diffusivity of the Graphene Composites

The performance of TIMs can be characterized by its thermal resistance, $R_{TIM}$, with specific bounding surfaces: $R_{TIM} \equiv H / K_A = H / K + R_{C1} + R_{C2}$. Here $H$ is the bond line thickness (BLT), $K$ is thermal conductivity of TIMs, $R_{C1}$ and $R_{C2}$ are the thermal contact resistances of the TIM layer with the two bonding surfaces. In this definition, $K_A$ is the effective or apparent thermal conductivity of the TIM with two contact resistances. The magnitudes of $R_{TIM}$ and $K_A$ depend on the thermal conductivity of the TIM, BLT and the thermal contact resistances, which are affected, in their turn, by the surface roughness, temperature and viscosity. The values of $R_{TIM}$ or $K_A$ have to be determined with practical BLT. Rewriting the above equation for $K_A$ one can get its dependence on BLT and contact resistances: $K_A = K \times \left[1 + (K / H)(R_{C1} + R_{C2})\right]^{-1}$. The apparent thermal conductivity, $K_A$, is a more practical metric for comparing TIM performance with actual bonding than the thermal conductivity, $K$, measured for bulk composite samples. For the proof-of-concept demonstration of the flake alignment approach we focus on determining $K_A$ of the composites with relevant connecting surfaces and temperature rise in actual devices. Detailed study of $R_{TIM}$ with various BLT and under different pressures is beyond the scope of this work.

The TIM composites were prepared with conventional curing epoxy and with non-curing commercial TIMs used for IC chip packaging. We first tested and verified that the apparent thermal conductivity of non-curing TIMs increases with the addition of LPE graphene fillers. The apparent thermal conductivity was measured by two different techniques. The first technique involved the TIM Tester (Analysis Tech 1400) that adheres to ASTM standards. The results of the measurements with the TIM Tester gave the apparent value of the thermal conductivity, which includes the thermal contact resistance with the connecting surfaces of interest, e.g. Cu to Cu or Si to aluminum (Al). The second technique used "laser flash" methods (Netzsch LFA 477 Nanoflash) compliant with the international standard ASTM E-1461, DIM EN 821 and DIN 30905. Following this technique, we measure the thermal diffusivity, which, in turn, is used to determine the thermal conductivity via the equation $K = \rho \alpha C_p$, where K is the thermal conductivity, $\rho$ is the mass density, $\alpha$ is the thermal diffusivity and $C_p$ is the specific heat. The Xenon flash lamp introduces an energy pulse to one side of the sample and the time dependent temperature rise is measured using an infrared (IR) detector (In-Sb) on the opposite side of the sample. All samples were custom fitted into planar circles with a 12.6 mm diameter in order to properly fit into the sample holder for cross-plane measurements. A washer-like mask was used to prevent any leakage of light from the Xenon lamp to the In-Sb detector on the opposite side of the sample.

Figure 4 shows representative results of the thermal conductivity measurements for the non-curing TIMs with graphene and few-layer-graphene fillers as the loading changes from zero to 6 wt. %. One can see that $K_A$ monotonically increases with $f$. The higher loadings were not practical owing to the increased viscosity and reduced uniformity of graphene dispersion. Addition of a small fraction ($f=6\%$) of randomly oriented graphene and FLG improved the apparent thermal conductivity of TIM spread between two Al plates by more than a factor of two. The results were typical for different measurements of non-curing TIMs. The increase in $K_A$ was achieved without optimization of the composition of the matrix material for additional graphene fillers. The apparent thermal conductivity revealed almost no dependence on temperature, which is characteristic for disordered materials and beneficial for practical applications.

[Figure 4: Thermal Data for Non-curing TIM with Random Graphene]

Figure 5 presents the thermal diffusivity data across a Cu-TIM-Cu "sandwich" measured using the "laser flash" technique. The mixture of the graphene and FLG fillers has been functionalized with the magnetic nanoparticles. In one sample the fillers were left randomized while in another they were oriented by placing the sample on a flat permanent magnet. The thermal diffusivity, $\alpha$, for conventional TIM without graphene was $\alpha \sim 0.23$ mm$^2$/s. This diffusivity value also includes the effect of the contact resistance with Cu plates. The addition of ~1 wt. % of graphene – FLG random fillers increases the apparent diffusivity by about a factor of 1.5. The increase is substantially larger – by a factor of 3.8 – for the oriented graphene fillers of the same loading fraction. For the realistic BLT, the fillers do not extend all the way from one surface to another and do not form a complete percolation network. However, alignment of the fillers along the direction of the heat flow substantially improves the heat conduction properties.

[Figure 5: Thermal Diffusivity]

In order to further elucidate the effect of alignment of graphene fillers we show the ratio of the apparent thermal conductivity in TIMs with the graphene fillers, $K_m$, to that in the reference TIM without the fillers, $K_o$ (see the Figure 6). These measurements were performed for TIMs with 1 wt. % of graphene inserted between two Cu films. As one can see the enhancement of the thermal conductivity is substantially higher for the TIMs with oriented graphene fillers rather than random fillers. However, even random graphene and FLG fillers at low loading increase the apparent thermal conductivity by a factor of 1.5. These data show that the benefits of the low loading of graphene fillers reported previously for bulk samples [19 – 20] are preserved for TIM layers with small BLT squeezed between two relevant surfaces. In order to increase the heat conduction properties, one can either increase the loading of graphene fillers, within certain limits, or use a filler alignment procedure to achieve the same enhancement with smaller loading.

[Figure 6: Thermal Conductivity Ratio]

The orientation of graphene fillers results in the enhanced heat conduction in composites with epoxy base as well. As one can see in Figure 7 that the enhancement of the apparent thermal conductivity with the oriented graphene fillers is twice as strong as that with the random fillers at small loading ($f = 1$ wt. %) of the graphene. The ratio of the apparent thermal conductivity of the composite to that of the base epoxy $K_m/K_o \sim 3.2$ for the oriented fillers and $K_m/K_o \sim 1.7$ for the random fillers. This result is obtained for apparent thermal conductivity, which includes the effect of the thermal contact resistances to the connecting metal surfaces. The enhancement for the bulk thermal conductivity can be stronger. It is interesting to note that the temperature dependence of the thermal conductivity and thermal diffusivity of composites with oriented fillers is somewhat different from that of composites with random fillers (see Figures 3, 4 and 5). The thermal conductivity of composites with random fillers increases with temperature, which is characteristic of amorphous and disordered materials [18–20, 56–58]. Similar weakly increasing trend was observed in other TIMs with random graphene fillers [18-21]. In contrast, the decrease in the thermal conductivity of composites with oriented graphene fillers can be explained by the decreasing viscosity at elevated temperatures, resulting in partial loss of the filler orientation.

[Figure 7: Thermal Conductivity of Epoxy]

We have also tested the performance of oriented graphene fillers with the thermal phase change materials (PCM) commonly used for thermal management of photovoltaic solar cells. The room temperature thermal conductivity of conventional PCM between two Cu plates was $K_o$=0.3 W/mK. The thermal conductivity of the composites with random and oriented graphene fillers was determined to be $K_m$=0.6 W/mK and $K_m$=1.25 W/mK, respectively. As with other base materials orientation allowed us to increase the enhancement by a factor of two at relatively low graphene loading fractions. The temperature dependence of the thermal conductivity was similar to that of non-curing and epoxy composites. In addition, we  verified that oriented graphene fillers perform better than the random graphene fillers, which were not functionalized with magnetic nanoparticles. Our results indicate that the filler orientation approach works with a wide variety of base materials used in passive thermal management.

The prior work with aligned fillers for thermal management applications was mostly focused on CNT arrays. The reported experiments have shown that a dense array of vertically aligned CNTs grown on Si and Cu substrates can provide thermal resistance values, which are less than 20 mm$^2$K/W [30–37]. The values obtained for CNT arrays are comparable to commercially available solders with the thermal resistances ranging from 7 to 28 mm$^2$K/W [30]. However, the reported vertically aligned CNT arrays were grown by CVD or PECVD techniques at temperatures of about 750°C. In the cases, when the lowest thermal resistance values were achieved, the CNT arrays had to be grown on both connecting surfaces. This is technologically a challenging and expensive approach. Our proposed technique is inexpensive, scalable and can be implemented with conventional equipment for TIM dispersion. The alignment only requires pads with embedded flat permanent magnets. One should note that the tested TIMs with self-aligning "magnetic graphene" fillers have not been optimized for loading fraction, flake size and viscosity of the matrix. However, the proof-of-concept measurements demonstrating filler alignment with magnetic field suggest that thermal resistances below 20 mm$^2$K/W are achievable with practical BLT values.

## IV.     Temperature Rise Testing in Computers

The thermal conductivity measured for bulk samples is a material metric, which does not completely characterize how well it will perform in practice. The apparent thermal conductivity measured for realistic BLT, which includes the effect of the thermal interface resistances, is a more informative characteristic of TIMs. However, the metric used in industry for assessment of the suitability of TIM for a specific application is a temperature rise in a given device or a system with specific TIM.   We conducted temperature rise, $\Delta T$, testing using a high-end desktop computer. These experiments allowed us to assess the TIM efficiency in transferring generated heat away from the computer processing unit (CPU). For these practical tests, the conventional TIM supplied with the CPU package was removed and replaced with our graphene-enhanced TIMs. To achieve alignment of the functionalized graphene fillers the CPU assemblies were placed on a flat permanent magnet. Below we describe the details of the temperature rise testing in CPUs of desktop computers.

To assess TIM efficiency in transferring generated heat away from CPU we assembled a custom desk-top computer system. The temperature rise measurements were performed on a CPU under the stress-test conditions to guarantee controlled constant-power output. The CPU was actively cooled with liquid that passed through channels beneath that copper heat-sink attached to the CPU, with further heat exchange in radiators. The active cooling setup allows for control of environmental variable that significantly affects the temperature of the CPU's core. Conventional air-cooled setups use ambient air, which has temperature widely fluctuating over the test time. The selected CPU (Intel® Core™ i7-4770K) was able to generate high thermal density power output while offering embedded reliable temperature monitoring capabilities. It draws 84 W at maximum power consumption during the thermally significant period while running stress-test software. Each of its four physical cores had an embedded thermistor for in-situ temperature monitoring. Averaging of the four readings was used to provide more reliable and noise-resistant temperature-rise data. The active temperature management was ensured with a liquid closed-loop cooling system (Corsair®). The cooling system offered high degree of control and thermal stability due to large radiator area and high heat capacity of circulating liquid. The round copper heat-sink disk was positioned on a maze of embedded water channels to maximize the effective thermal intake area. A large surface area facilitated coverage of the entire CPU's secondary heat sink. The polished surface offers better heat sink-to-TIM bonding and reduced the number and depth of the surface trenches prone to trapping thermally-insulating air. The CPU was kept at its peak computing capacity and thus generating the maximum possible thermal density using three stress-test software tools: InterBurnTest, OCCT and LINX. The software overloaded CPU with unending series of floating-point arithmetic operations and numeric linear algebra computations for user-specified period of time. To monitor and log the thermal state of all physical cores, two temperature acquisition programs were used: CoreTemp and RealTemp. Both are capable of reading raw data from four embedded thermistors at specified time intervals, buffer the data, average to eliminate any erroneous fluctuations in readings and log data to a text file. During the measurements, RealTemp performed logging at 3 second intervals and CoreTemp performed 1000 ms thermistor polling and averaging over 10 readings with 10 second single-value logging. All experimental runs were conducted over the period of 24 hours at 100% CPU loading. The profile curve for commercial (IceFusion) TIMs served as a base for comparison of efficiency of the synthesized TIMs with oriented graphene fillers. The decrease in the height of temperature profile with respect to established baseline indicated lower temperatures in the cores of the CPU. BLT was controlled with a micrometer and maintained approximately constant for various TIMs. The "magnetic graphene" filler alignment was achieved by placing CPU heat sinks on flat permanent magnets (N52 Neodymium), which provided ~1.5 T field. After the magnetic alignment the CPU with TIM was assembled with the liquid-cooled heat sink for stress-testing and temperature-rise monitoring. The testing facility was equipped with a climate control. The ambient temperature fluctuations were negligible (~1.5 ℃ drift within 24 hours).

In Figure 8 we show the temperature rise as a function of time inside a computer operating with a heavy computational load. The data were obtained using two temperature acquisition programs (see METHODS section). The data in Figure 8 (a) is for TIMs with random graphene fillers. For proper comparison the graphene fillers have also been functionalized with magnetic nanoparticles but not oriented with the magnetic field. One can see that the increasing weight fraction, $f$, of graphene fillers results in decreasing CPU temperature. At $f \approx 4\%$ the temperature rise $\Delta T=55^{\circ}C$, which is lower by $10^{\circ}C$ of $\Delta T$ recorded for the reference commercial TIM. The performance of TIMs with 1% of graphene does not differ substantially from the reference TIM. Approximately $5^{\circ}C$ reduction in $\Delta T$ was achieved with $f \approx 2\%$ of random graphene fillers. Orientation of the functionalized graphene fillers results in drastic improvement of the TIM efficiency. Figure 8 (b) shows that the temperature rise is reduced in the CPU by as much as $10^{\circ}C$ after 15 hours of operation when TIM with $f \approx 1\%$ of the oriented graphene fillers is used. This reduction is substantial for practical applications. In some device technologies the reduction of the temperature rise by $20^{\circ}C$ translates into an order of magnitude increase of the device life-time [7, 10].

[Figure 8: Temperature Rise Tests]

## V.     Conclusions

We conducted proof-of-concept investigation of the thermal interface materials with self-aligning "magnetic graphene" fillers. Functionalization of LPE graphene flakes with $Fe_3O_4$ nanoparticles allowed us to align graphene fillers in an external magnetic field during dispersion of the thermal paste to the connecting surfaces. The graphene filler alignment resulted in a strong increase of the thermal conductivity of the composites. The self-aligning "magnetic graphene" fillers improve heat conduction in composites with both curing and non-curing matrix materials. The testing conducted with computer chips demonstrated the temperature rise decrease by as much as 10°C with use of the non-curing thermal interface material with ~1 wt. % of the oriented graphene fillers. The demonstrated TIMs with self-aligning "magnetic graphene" fillers present a less expensive alternative to CNT arrays grown by CVD method at high temperature. In our design, the aligning of graphene fillers can be accomplished during TIM dispersion to the connecting surfaces in regular industrial environment. The obtained results suggest that TIMs with functionalized graphene fillers have a potential for a major development in thermal management of advanced electronics.

## Acknowledgements

The work at Quantum Seed LLC was supported via the National Science Foundation (NSF) project IIP-1345296 SBIR: Ultra-High-Efficiency Thermal Interface Materials Based on Self-Aligned Graphene Fillers (submitted June 2013, funded November 2013). The work at UC Riverside was supported, in part, via the University of California Office of the President (UCOP) grant 268947: Graphene Based Thermal Interface Materials and Heat Spreaders (funded August 2013), NSF project CMMI-1404967 Hybrid Computational Experimental Engineering of Defects in Designer Materials.

## Authors Contributions

A.A.B. conceived the idea of the magnetically aligned graphene fillers (University of California invention disclosures UC 2013-156, UC 2013-812, UC 2014-246 submitted in 2013; patent filed in 2014), conducted data analysis and wrote the manuscript; J.R. conducted materials synthesis and characterization, and contributed to the manuscript preparation; S.L. performed temperature rise tests; R.S., S.R., F.K. and M.S. carried out materials synthesis and thermal measurements with a range of thermal paste composites; M.P.B. assisted with materials preparation and data processing.

**REFERENCES**

**Figure Captions**

**Figure 1:** Schematic of the technique for graphene and few-layer-graphene functionalization with magnetic nanoparticles. The steps include: addition of PSS to graphene solution resulting in graphene surface coating with the "primer"; addition of PDDA, which sticks to PSS "primer" via electrostatic interactions and provides distribution of positive charges on graphene fillers; addition of the solution of magnetic nanoparticles, which attach to PDDA layer via electrostatic interaction during the stirring; mixing of the magnetically functionalized graphene fillers with the matrix material resulting in the composite, which is ready for filler alignment with an external magnetic field.

**Figure 2:** Illustration of the technological steps for synthesis of thermal interface materials with graphene and few-layer-graphene fillers. The liquid phase exfoliation of graphene starts with raw graphite source material. The functionalization step involves attachment of magnetic nanoparticles shown in the previous figure. The entire process is scalable and less expensive than CVD growth of ordered arrays of carbon nanotubes.

**Figure 3:** Preparation and characterization of the graphene and few-layer-graphene fillers functionalized with magnetic nanoparticles. (a) Scanning electron microscopy of graphene and few-layer-graphene flakes synthesized by the liquid-phase exfoliation technique. (b) Transmission electron microscopy image of the graphene flake with attached $Fe_3O_4$ nanoparticles. Observed agglomeration of graphene flakes did not prevent alignment and thermal applications. (c) Photograph illustrating a reaction of the magnetically functionalized graphene fillers on a permanent magnet (B=1.5 T). (d) Two copper foils with the functionalized graphene TIM between them placed on a flat permanent magnet for alignment of the "magnetic graphene" fillers. (e)

Optical microscopy image of epoxy with aligned graphene fillers. Higher loading of graphene and few-layer-graphene fillers were used to reveal alignment at a larger length scale.

**Figure 4:** Apparent thermal conductivity of a representative commercial TIM with different loading fraction, $f$, of graphene without alignment. The apparent thermal conductivity, which includes the thermal boundary resistance (TBR) with connecting surfaces, monotonically increases with $f$ in the examined range. The higher loadings were not practical owing to the increased viscosity and reduced uniformity of graphene dispersion.

**Figure 5:** Apparent thermal diffusivity for Cu-TIM-Cu structure as a function of temperature. The data are shown for a sample where the graphene fillers were left random and another sample where they were oriented with a help of a flat permanent magnet. Note that the heat dissipation via TIM with the oriented graphene fillers is substantially better than in the reference commercial TIM and TIM with random graphene fillers.

**Figure 6:** Ratio of the apparent thermal conductivity of the graphene-enhanced TIM to that of the reference TIM as a function of temperature. The data are shown for TIMs with random and oriented graphene fillers. Note that the enhancement of thermal conductivity is substantially higher for TIMs with the oriented graphene fillers.

**Figure 7:** Apparent thermal conductivity in epoxy composite with the random and oriented graphene fillers. The data are shown for 1% of graphene loading. Note that the enhancement is twice as strong for the oriented fillers at this low loading fraction.

**Figure 8:** Temperature rise inside a computer CPU as a function of time. The data are presented for CPU packages that utilized TIMs with random graphene fillers (a) and oriented graphene fillers (b). As a reference the temperature rise in CPU package with conventional commercial TIM is shown with a black line. The insets show the backside of a computer chip with applied TIM.  The data were obtained using two different temperature acquisition programs.

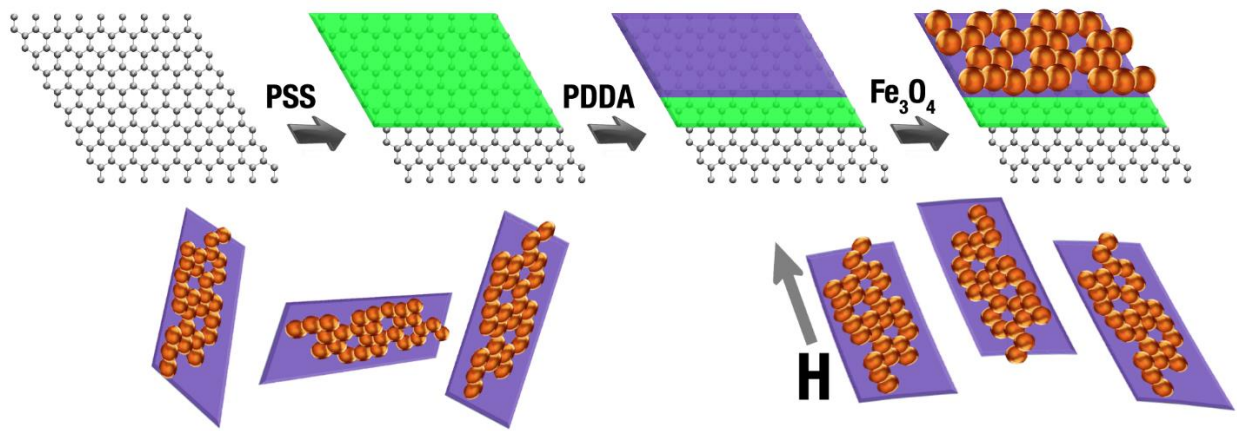

Figure 1

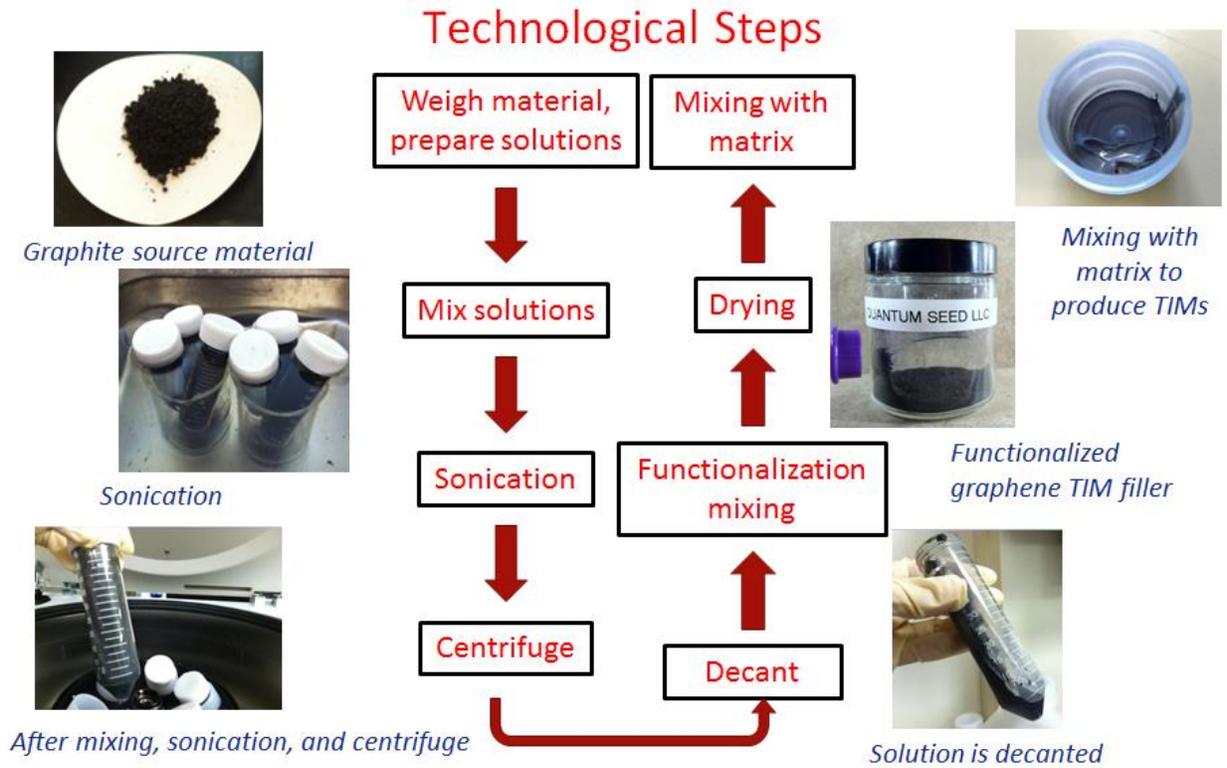

Figure 2

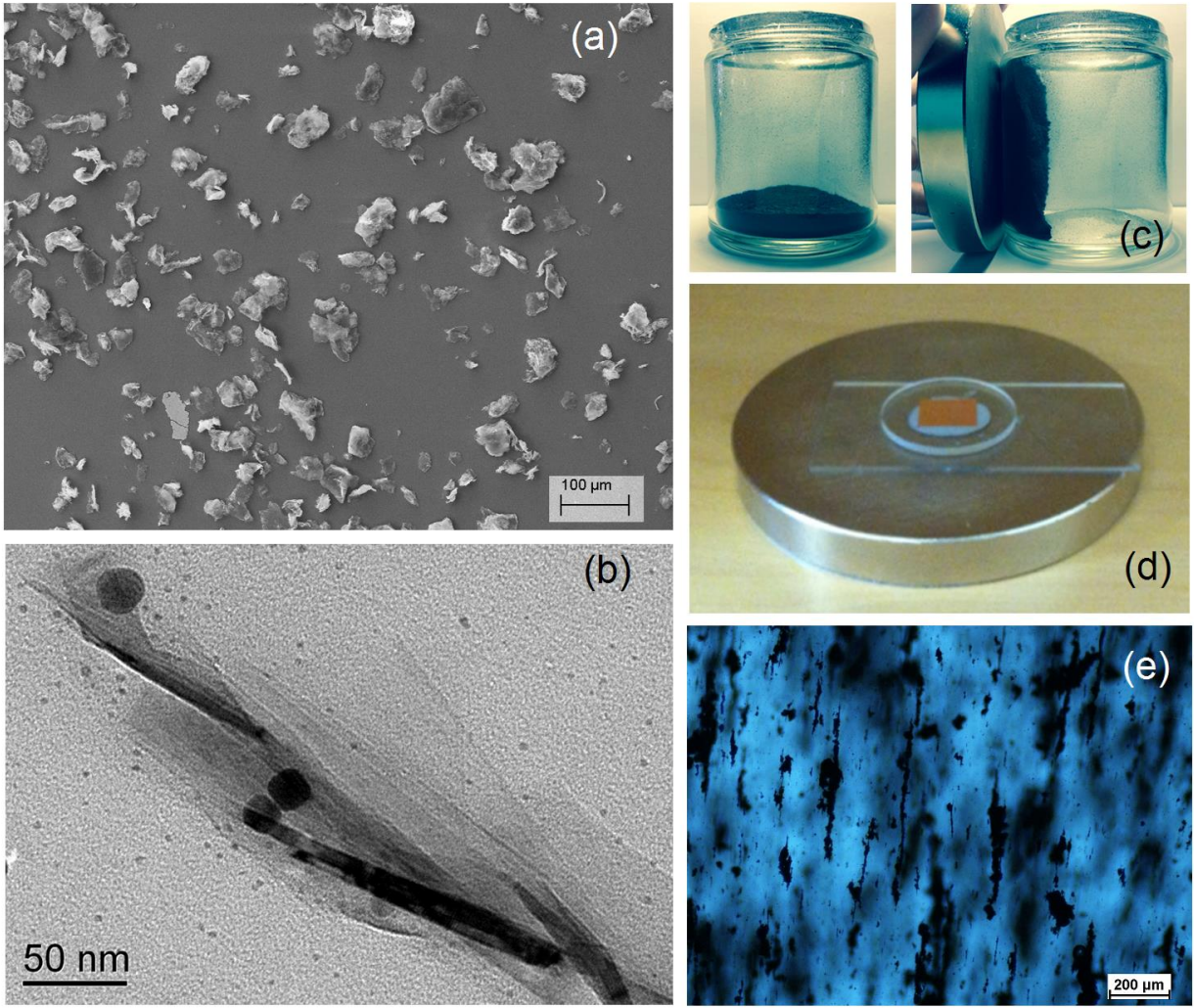

Figure 3

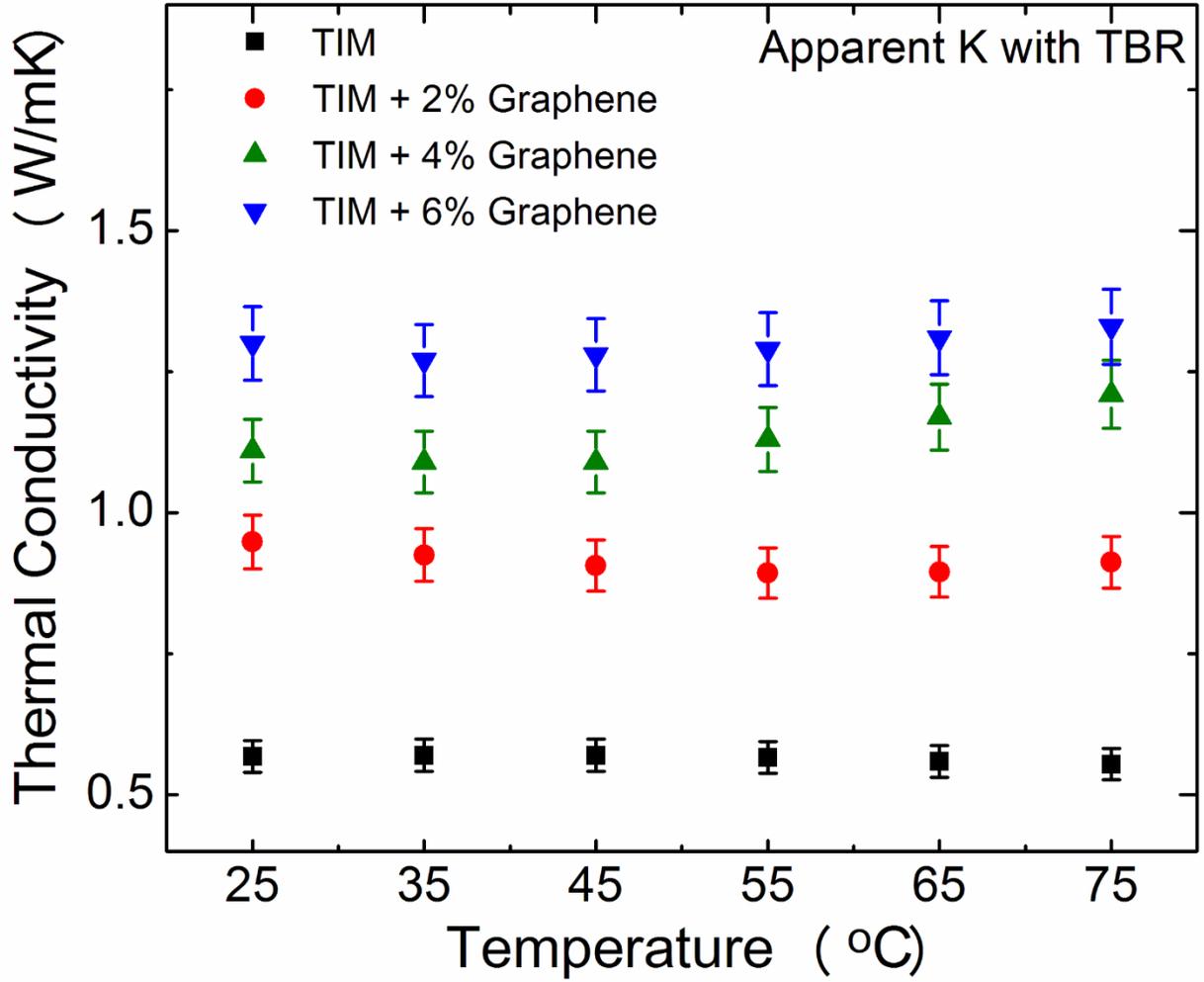

Figure 4

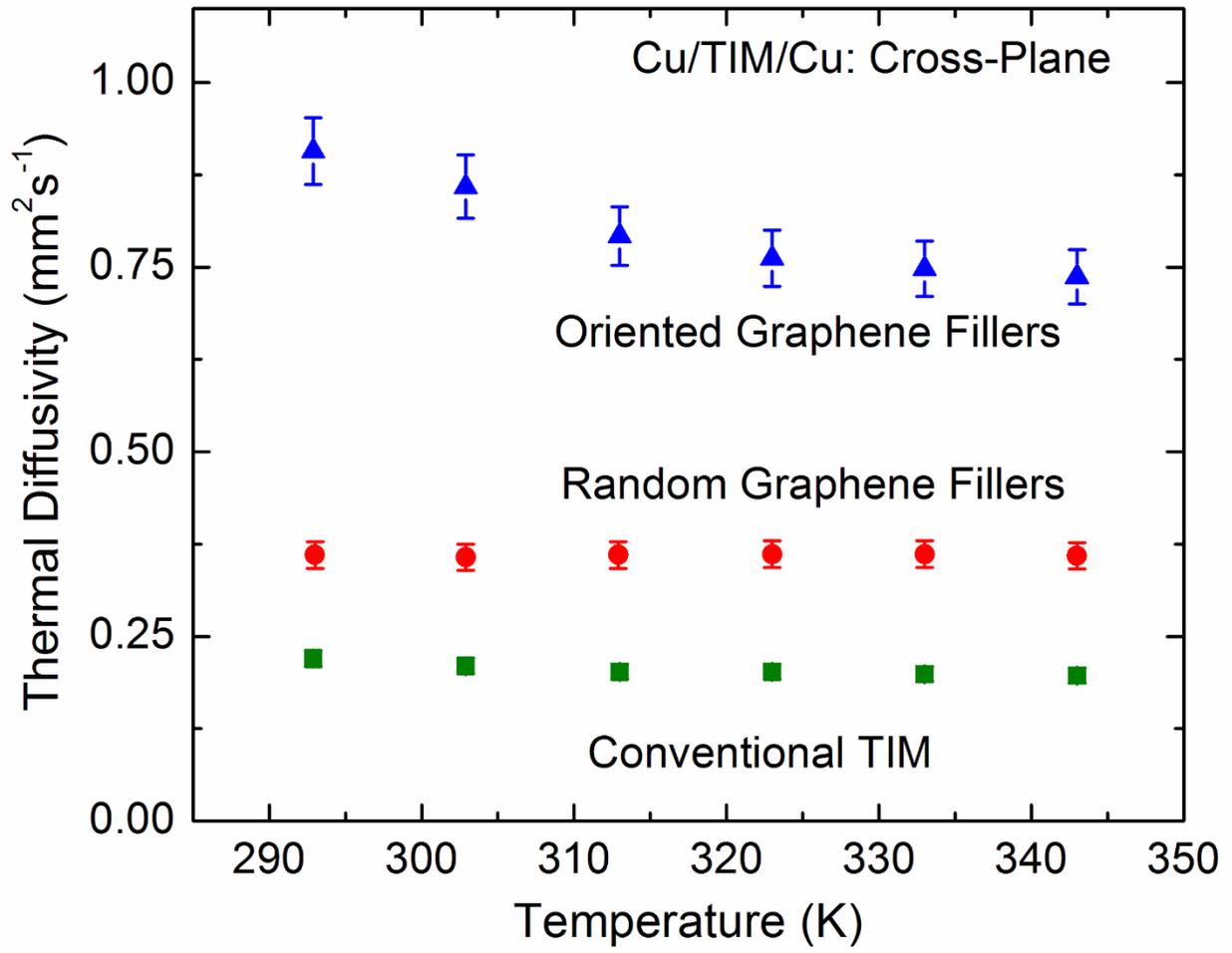

Figure 5

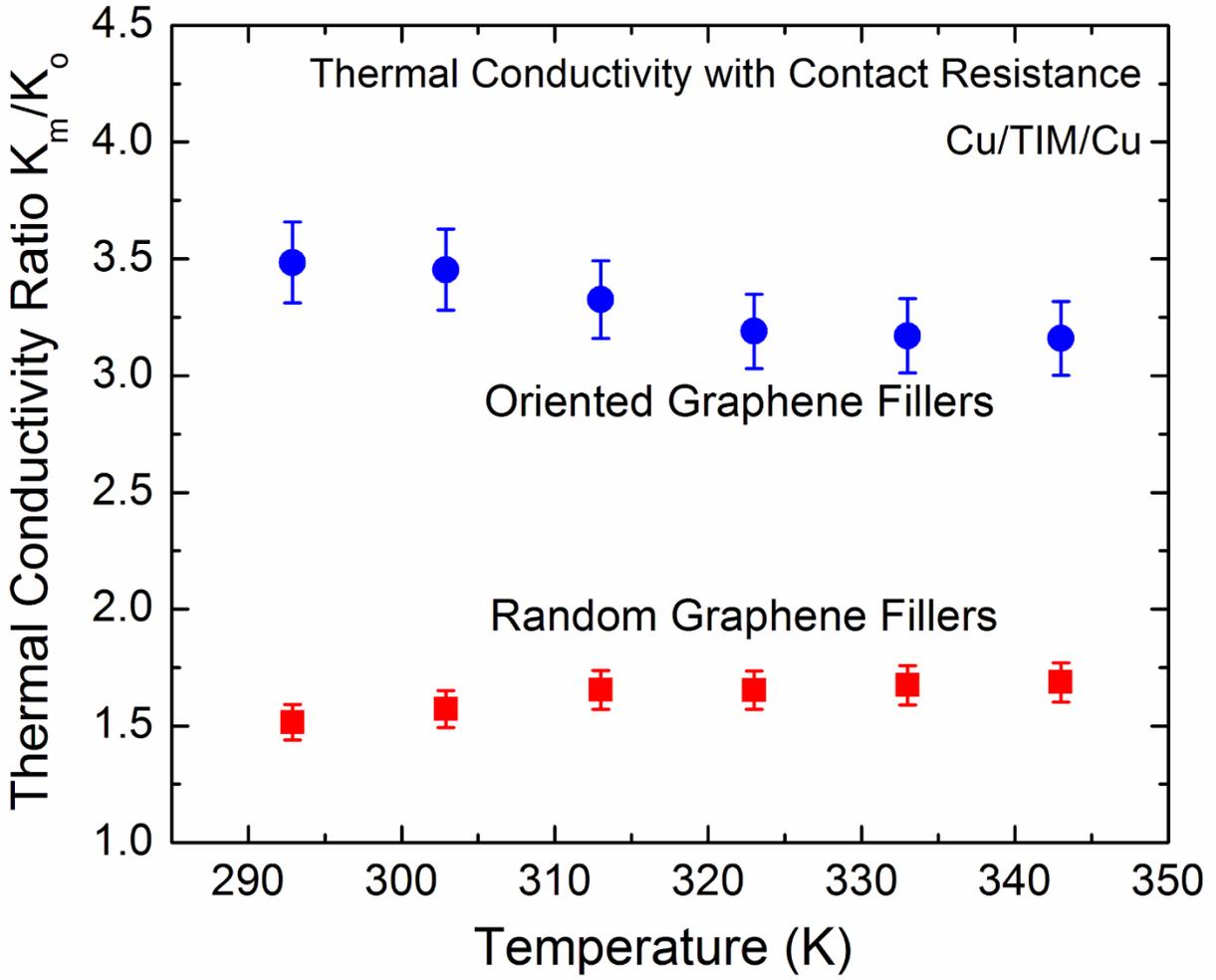

Figure 6

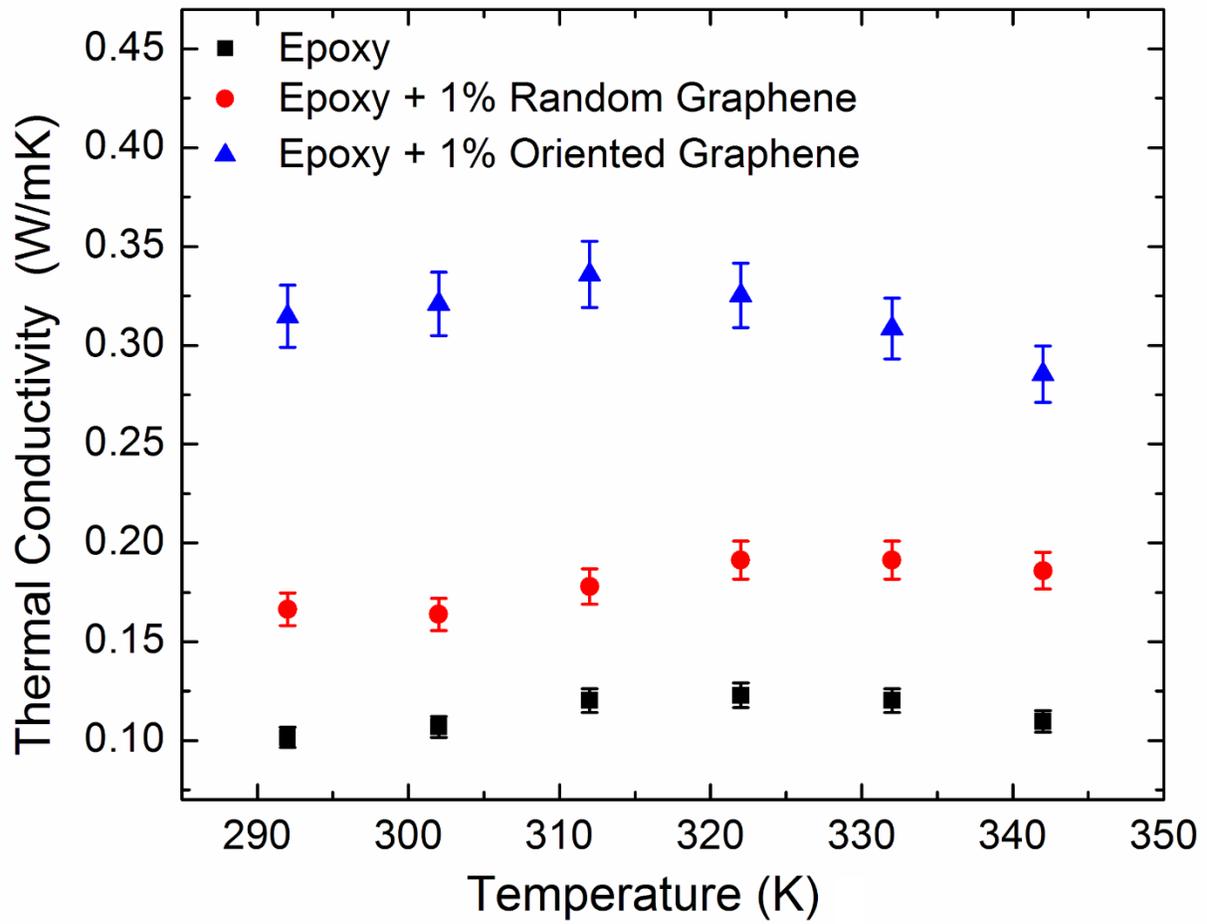

Figure 7

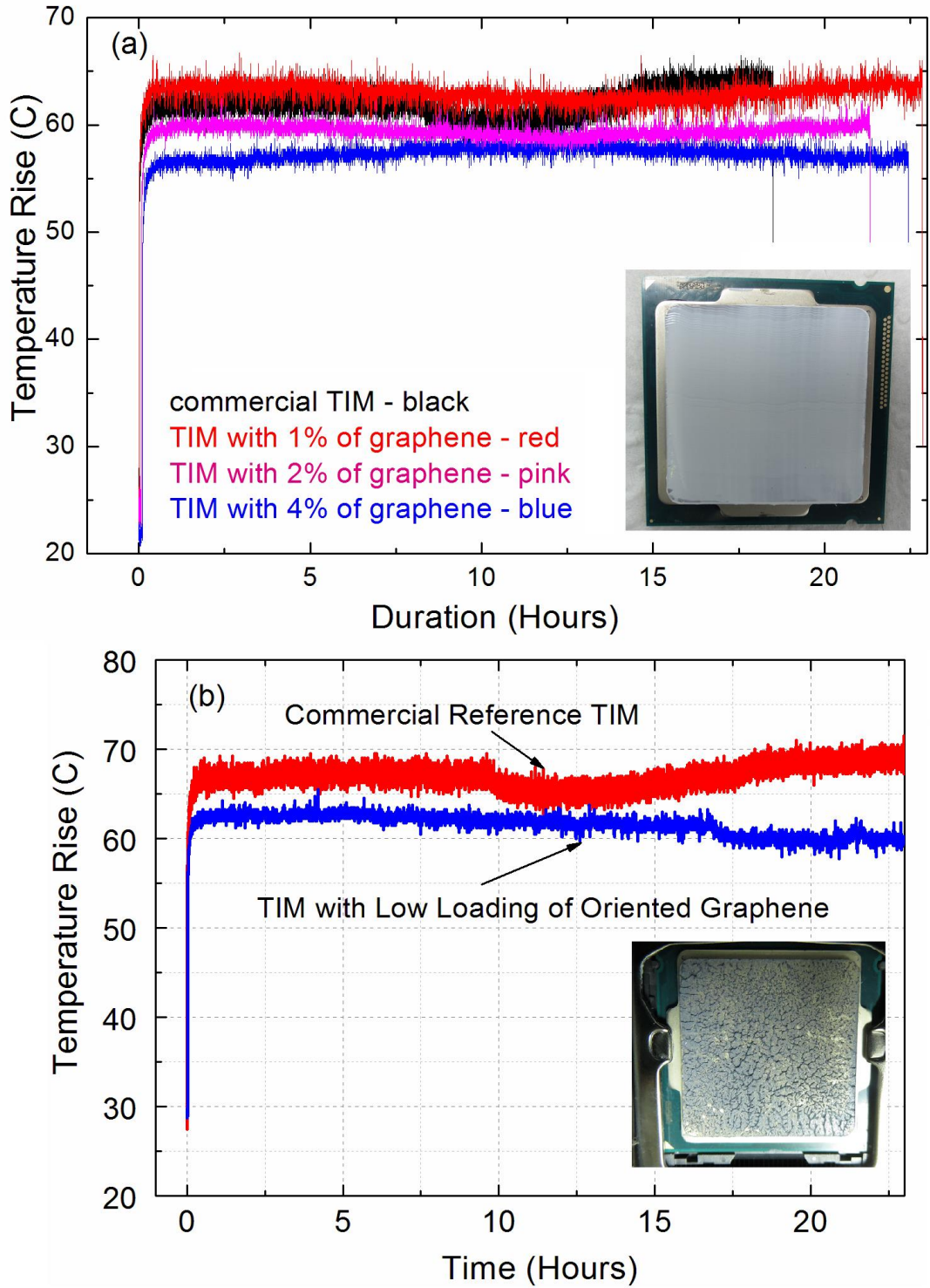

Figure 8